\documentclass[reprint, 10pt, amsmath, amssymb, floatfix, aps, prd, superscriptaddress, nofootinbib, nobibnotes,twocolumn]{revtex4-1}

\usepackage{orcidlink}
\usepackage{acro}
\usepackage{braket}

\DeclareAcronym{gb}{
    short = {GB},
    long = {Gauss-Bonnet},
}
\DeclareAcronym{sgb}{
    short = {SGB},
    long = {scalar Gauss-Bonnet},
}
\DeclareAcronym{twa}{
    short = {TWA},
    long = {truncated Wigner approximation},
}
\DeclareAcronym{usr}{
    short = {USR},
    long = {ultra slow-roll},
}

\DeclareAcronym{GW}{
  short = GW ,
  long = gravitational wave ,
  short-plural = s
}
\DeclareAcronym{LIGO}{
  short = LIGO ,
  long = Laser Interferometer Gravitational-wave Observatory ,
  short-plural =
}
\DeclareAcronym{LISA}{
  short = LISA ,
  long = Laser Interferometer Space Antenna ,
  short-plural =
}
\DeclareAcronym{SKA}{
  short = SKA ,
  long = Square Kilometre Array ,
  short-plural =
}

\DeclareAcronym{SNR}{
	short = SNR ,
	long = signal-to-noise ratio ,
	short-plural =
}

\DeclareAcronym{PTA}{
	short = PTA ,
	long = pulsar timing array ,
	short-plural =
}

\DeclareAcronym{FLRW}{
  short = FLRW ,
  long = Friedmann-Lemaitre-Robertson-Walker ,
  short-plural =
}

\DeclareAcronym{SIGW}{
	short = SIGW ,
	long = scalar induced gravitational wave ,
	short-plural =  s
}

\DeclareAcronym{PBH}{
	short = PBH ,
	long = primordial black hole ,
	short-plural =  s
}

\DeclareAcronym{SMBHB}{
  short = SMBHB ,
  long = supermassive black hole binary ,
  short-plural = s
}

\DeclareAcronym{1PI}{
	short = 1PI ,
	long = one-particle irreducible  ,
	short-plural =
}

\DeclareAcronym{1PR}{
	short = 1PR ,
	long = one-particle reducible  ,
	short-plural =
}

\DeclareAcronym{KDE}{
  short = KDE ,
  long = kernel density estimator ,
  short-plural = s
}

\DeclareAcronym{BPBHM}{
  short = BPBHM ,
  long = binary primordial black hole merger,
  short-plural = s
}

\DeclareAcronym{CMB}{
	short = CMB ,
	long = cosmic microwave background ,
	short-plural =
}
\DeclareAcronym{DM}{
	short = DM ,
	long = dark matter ,
	short-plural =
}

\DeclareAcronym{BBN}{
	short = BBN ,
	long = Big-Bang nucleosynthesis ,
	short-plural =
}

\DeclareAcronym{LN}{
	short = LN ,
	long = log-normal  ,
	short-plural =
}

\DeclareAcronym{BPL}{
	short = BPL ,
	long = broken power-law ,
	short-plural =
}

\DeclareAcronym{SGWB}{
	short = SGWB ,
	long = stochastic gravitational	wave background ,
	short-plural =  s
}

\DeclareAcronym{LSS}{
	short = LSS ,
	long = large scale structure ,
	short-plural =
}

\DeclareAcronym{RD}{
	short = RD ,
	long = radiation-dominated ,
	short-plural =
}

\DeclareAcronym{PLS}{
	short = PLS ,
	long = power low sensitivity ,
	short-plural =
}

\DeclareAcronym{MAP}{
	short = MAP ,
	long = maximum a posterior ,
	short-plural =
}

\DeclareAcronym{PGW}{
	short = PGW ,
	long = primordial gravitational wave ,
	short-plural =  s
}

\DeclareAcronym{BAO}{
	short = BAO ,
	long = baryon acoustic oscillations ,
	short-plural =
}

\DeclareAcronym{TSIGW}{
	short = TSIGW ,
	long = tensor-scalar induced gravitational wave ,
	short-plural =  s
}

\def\dd{{\mathrm{d}}}

\begin{document}

\author{Fei-Yu Chen\orcidlink{0000-0002-8674-9316}}
\email{chenfy@tju.edu.cn}
\affiliation{College of Physics and Materials Science, Tianjin Normal University, Tianjin 300387, China}

\author{Jing-Zhi Zhou\orcidlink{0000-0003-2792-3182}}
\email{zhoujingzhi@hau.edu.cn}
\affiliation{Department of Mathematics and Physics, Huaian University, Meicheng East Road 1, Huaian, Jiangsu 223200, China}

\author{Zhi-Chao Li\orcidlink{0009-0005-7984-2626}}
\affiliation{Center for Joint Quantum Studies and Department of Physics,
School of Science, Tianjin University, Tianjin 300350, China}

\author{Di Wu\orcidlink{0000-0001-7309-574X}}
\affiliation{School of Fundamental Physics and Mathematical Sciences, Hangzhou Institute for Advanced Study, University of Chinese Academy of Sciences, Hangzhou 310024, China}

\begin{abstract}
	We use lattice methods to study inflation in the \ac{sgb} gravity theory. We focus on the ultra-slow-roll scenario with the peak frequency falls into the PTA band. In the parameter range we consider, we find that the lattice results exceed the perturbative predictions, which differs from the case in Einstein gravity. We further find that lattice corrections become significant when the peak of the primordial curvature spectrum reaches  $\sim 10^{-2}$. We calculate the energy density spectra of second-order \acp{SIGW} using the primordial power spectra obtained from both the lattice method and the traditional perturbative method, and analyze the \ac{sgb} model in light of current \ac{PTA} observations. Our results indicate that lattice corrections enhance the ability of the \ac{sgb} model to dominate the \ac{PTA} observations.
\end{abstract}
\title{Inflation on the lattice: scalar Gauss-Bonnet single field inflation}

\maketitle

\section{Introduction}
\label{sec:intro}
Cosmological inflation is currently a highly successful theory describing the very early evolution of the universe. Within the standard cosmological framework based on the \ac{FLRW} metric, inflation predicts a nearly scale-invariant primordial power spectrum of curvature perturbations, which is consistent with observations on \ac{CMB} scales \cite{Planck:2018vyg}. Moreover, experimental observations have placed strong constraints on the parameters of the power spectrum at these scales. However, the constraints from observations are much weaker on smaller scales. Large‑amplitude primordial perturbations on small scales can lead to the formation of \acp{PBH} \cite{Kawai:2021edk,Carr:2020xqk,Saito:2008jc,Garcia-Bellido:2017aan,Bartolo:2018evs,Ferrante:2022mui,LISACosmologyWorkingGroup:2023njw,Iovino:2024tyg,DeLuca:2020agl,Choudhury:2023rks,Fang:2025vhi,Pi:2026uxd,Allegrini:2026jqt,Wang:2026sgz,Bi:2026zlt}. In addition, primordial curvature perturbations can serve as source terms for second‑order tensor perturbations, generating \acp{SIGW} that contribute to the \ac{SGWB}, offering a potential explanation for \ac{PTA} observations \cite{Domenech:2021ztg,Chang:2023aba,Cai:2018dig,Baumann:2007zm,Ananda:2006af,NANOGrav:2023hvm,Franciolini:2023pbf,Vaskonen:2020lbd,Wang:2023ost,Domenech:2025ccu,LISACosmologyWorkingGroup:2025vdz,LISACosmologyWorkingGroup:2024hsc,Chen:2026scy,Bhaumik:2026rzf,Caporali:2026qhc,Zhang:2026hqg,Zhao:2026jax,Domenech:2026nun}. In this work, we focus on inflationary models capable of producing large primordial curvature perturbations on small scales.

The calculation of the primordial power spectrum is typically performed at linear order \cite{Karam:2022nym,Zhou:2020kkf,Ragavendra:2020sop,Ballesteros:2020qam,Cai:2019bmk,Ozsoy:2019lyy,Chen:2019zza,Carrilho:2019oqg,Pi:2021dft,Guzzetti:2016mkm,Malik:2008im,Di:2017ndc,Kim:2025dyi,Huang:2026bwa,Rashidi:2026wtj,Dalianis:2026tty}. However, when curvature perturbations become sufficiently large, higher-order corrections can no longer be neglected and may even lead to the breakdown of perturbation theory \cite{Davies:2023hhn,Firouzjahi:2023bkt,Braglia:2025cee,Fumagalli:2023loc,Cheng:2023ikq,Firouzjahi:2023btw,Firouzjahi:2023aum,Firouzjahi:2024sce,Fumagalli:2024jzz,Kawaguchi:2024rsv,Choudhury:2024dzw,Choudhury:2024dei}. The lattice method offers a potential solution to this issue \cite{Saha:2026cay,Launay:2025lnc,Jamieson:2025ngu}. The fundamental idea of the lattice method is to discretize spacetime and solve the full cosmological evolution equations, thereby non-perturbatively deriving the primordial power spectra. Ref.~\cite{Caravano:2024moy} provides a detailed analysis of the single-field ultra-slow-roll (USR) scenario in Einstein gravity and finds that when the peak of the linear primordial power spectrum reaches $10^{-3} \sim 10^{-2}$, noticeable deviations emerge compared to the results from lattice calculations.

Most single-field inflationary models considered in the literature are based on Einstein gravity. However, Einstein gravity might not be a complete theory of gravity, making it meaningful to explore modified theories of gravity. For instance, in the low-energy effective theory of string theory, higher-order curvature correction terms appear in the Lagrangian. The \ac{gb} term, ${\mathcal L}_{\rm GB} = R^2 - 4 R_{\mu\nu}R^{\mu\nu} + R_{\mu\nu\rho\sigma}R^{\mu\nu\rho\sigma}$, is a well-behaved quadratic curvature term because it yields second-order equations of motion. Refs.\cite{Kawai:2021bye,Kawai:2021edk} first investigated an inflationary theory of \ac{sgb} gravity, which involves a coupling term between a scalar field $\phi$ and the \ac{gb} term. This theory features a non-trivial fixed point of $\phi$, near which ultra-slow-roll inflation occurs, amplifying the primordial power spectrum within a certain frequency range. For the parameter set considered in that study, the peak of the primordial power spectrum reaches $10^{-2}$, which already falls within the range $10^{-3} \sim 10^{-2}$ where the linear approximation breaks down, as mentioned earlier. Although this range was originally derived from Einstein gravity and may not apply to \ac{sgb} gravity, it is still essential to investigate \ac{sgb} gravity and examine the results it yields. Therefore, in this paper, we consider the \ac{sgb} gravity theory and employ the lattice method to compute the primordial power spectrum, comparing the results with those obtained from perturbation theory.

This paper is organized as follows. In Sec. \ref{sec:gb-gravity} we review the \ac{sgb} gravity theory and the corresponding inflation model. In Sec. \ref{sec:lattice-method} we review the lattice approach to cosmic inflation, specifically, the lattice calculation of primordial scalar power spectrum. In Sec. \ref{sec:results} we present our results and discuss the implications. In Sec.~\ref{sec:SIGW}, we use the results obtained in the previous sections to compute the corresponding  energy density spectrum of \acp{SIGW} and examine its impact on current \ac{PTA} observations.  Finally, we conclude this work in Sec. \ref{sec:conclusion}. We set $c = \kappa = 1$ in this paper.

\section{SGB inflation theory}
\label{sec:gb-gravity}
The Lagrangian of \ac{sgb} gravity theory is given by
\begin{equation}
    \mathcal L = \frac{1}{2}R - \frac12\partial_\mu\phi\partial^\mu\phi - V(\phi) - \frac{\xi(\phi)}{16}\mathcal{L}_{\rm GB} \ ,
\end{equation}
where $\mathcal{L}_{\rm GB} = R^2 - 4R_{\mu\nu}R^{\mu\nu} + R_{\mu\nu\rho\sigma}R^{\mu\nu\rho\sigma}$, $V(\phi)$ and $\xi(\phi)$ are two functions at our choice. Assuming the background spacetime to be the FLRW one
\begin{equation}
    \dd s^2 = -\dd t^2 + a(t)^2(\dd x^2 + \dd y^2 + \dd z^2) \ ,
\end{equation}
the background equation of motion is given by
\begin{gather}
    3H^2 = \frac12 \dot\phi^2 + V(\phi) + \frac32 H^3\xi'(\phi)\dot\phi \ ,\nonumber\\
    \ddot\phi + 3H\dot\phi + \frac{\partial V_{\rm eff}}{\partial\phi}(\phi, H) = 0 \ ,\label{eq:flrw-eom-2}
\end{gather}
where $\dot{\phi} = \dd\phi/\dd t$, and the scalar effective potential $V_{\rm eff}$ is given by
\begin{equation}
	V_{\rm eff}(\phi, H) = V(\phi) + \frac32 H^2\left(\dot H + H^2\right)\xi(\phi) \ .
\end{equation}
The non-trivial fixed point $\phi_\ast$ satisfies
\begin{equation}
    V'(\phi) + \frac{1}{6}\xi'(\phi) = 0 \ ,
\end{equation}
which follows from $\dot H = \dot\phi = 0$.

Following Ref.~\cite{Kawai:2021edk}, we choose $\xi(\phi)$ to be a step-like function
\begin{equation}
    \xi(\phi) = \alpha\tanh[\xi_1(\phi - \phi_c)] \ ,
\end{equation}
and the scalar potential $V(\phi)$ is chosen to be the natural inflation model
\begin{equation}
    V(\phi) = \Lambda^4\left(1 + \cos\frac{\phi}{f}\right) \ ,
\end{equation}
where $\alpha$, $\xi_1$, $\phi_c$, $\Lambda$, $f$ are parameters, and $\phi_c$ is chosen to be the non-trivial fixed point, \textit{i.e.} $\phi_c = \phi_\ast$. Near the non-trivial fixed point, Eq. \eqref{eq:flrw-eom-2} becomes $\ddot\phi + 3H\dot\phi \approx 0$, which indicates an ultra-slow-roll regime and would result in an enhancement in the curvature power spectrum.

The quadratic action of the curvature perturbation $\zeta$ is given by
\begin{align}
    S_\zeta^{(2)} & = \frac12\int\dd t\dd^3x\frac{A_\zeta^2}{a}\left(a^2\dot\zeta^2 - C_\zeta^2\partial_i\zeta\partial^i\zeta\right)\nonumber\\
    & = \frac12\int\dd\eta\dd^3x A_\zeta^2 \left(\zeta'^2 - C_\zeta^2\partial_i\zeta\partial^i\zeta\right) \ ,
\end{align}
where ${}'$ denotes derivative with respect to the conformal time $\eta$. Coefficients $C_\zeta$ and $A_\zeta$ can be expressed in terms of background quantities $a$, $\phi$ and their time derivatives. For details, including the explicit expressions of $C_\zeta$ and $A_\zeta$, see Refs. \cite{Kawai:2021bye,Kawai:2021edk}.

Define the Mukhanov-Sasaki variable $\hat u = A_\zeta\zeta$ and perform a mode expansion
\begin{equation}\label{eq:u-mode-expansion}
	\hat u(\vec x, t) = \int\frac{\dd^3k}{\sqrt{(2\pi)^3}}\left[a_{\vec k} u_k(t) + a_{-\vec k}^\dagger u_k^\ast(t)\right] e^{i\vec k\cdot\vec x} \ ,
\end{equation}
where $a_k$ are the ladder operators obeying $[a_k, a_{k'}^\dagger] = \delta^{(3)}(\vec k - \vec k')$. $u_k$ is the mode function, and its equation of motion is given by
\begin{equation}\label{eq:curvature-pert-eom}
    u_k'' + \left(C_\zeta^2k^2 - \frac{A_\zeta''}{A_\zeta}\right)u_k = 0 \ , \  \qquad u_k = A_\zeta\zeta_k  \ .
\end{equation}
In the sub-horizon limit, the initial condition of the curvature perturbation is chosen as the Bunch-Davies vacuum, in which case the mode function takes the following asymptotic form
\begin{equation}\label{eq:scalar-bd-vacuum}
	u_k^{\rm BD} = \frac{1}{\sqrt{2C_\zeta k}}e^{-iC_\zeta k\eta} \ .
\end{equation}
The curvature power spectrum is given by
\begin{equation}
	\mathcal P_\zeta = \frac{k^3}{2\pi^2}\left|\frac{u_k}{A_\zeta}\right|^2  \ .
\end{equation}
In this perturbative approach, we first solve the background equation of motion Eq.\eqref{eq:flrw-eom-2} numerically. Then we solve the perturbation equation of motion \eqref{eq:curvature-pert-eom} numerically by initializing $u_k$ using \eqref{eq:scalar-bd-vacuum} at the time when the mode is deep inside the Hubble sphere, specifically we require $k / (aH) \sim 10^3$. We then evaluate $u_k$ until the mode exits the Hubble sphere where $k / (aH) \sim 10^{-3}$ and the mode function $\zeta_k$ approaches a constant.

\section{Lattice methodology}
\label{sec:lattice-method}
To analyse the nonlinear aspects of \ac{usr} in \ac{sgb} inflation theory, we use the lattice simulation method of inflation developed in \cite{Caravano:2021pgc,Caravano:2022epk}. In this section, we briefly review this method.

\subsection{Equation of motion}
We begin by discretizing the space as a cubic box of $N^3$ lattice points, with lattice spacing $\Delta x = L / N$, where $L$ is the physical edge length of the box. The inflaton field $\phi$ is defined by its values on each lattice site
\begin{equation}
	\phi(\vec x, t) \to \phi(\vec n L / N, t) \ ,
\end{equation}
where $\vec n = (n_x, n_y, n_z)$ is a vector containing the lattice coordinates. The momentum space is given by the reciprocal lattice
\begin{equation}
	\vec k = \frac{2\pi\vec n}{L} \ .
\end{equation}

In the lattice method, fields are evolved using classical equations of motion. The full equation of motion of $\phi$ has one extra term compared to \eqref{eq:flrw-eom-2}
\begin{equation}
	\ddot\phi + 3H\dot\phi + \frac{\partial V_{\rm eff}}{\partial\phi}(\phi, H) - \frac{\nabla^2\phi}{a^2} = 0 \ .
\end{equation}
We choose the discretized Laplacian to be
\begin{align}
	\nabla^2\phi(\vec n, t) & = \frac{1}{\Delta x^2}\sum_i \left[\phi(\vec n + \vec n_i, t) + \phi(\vec n - \vec n_i, t)\right.\nonumber\\
    & \qquad \left.- 2\phi(\vec n, t)\right] \ ,
\end{align}
where $\vec n_i$ is the direction vector at the $i$-th coordinate. This gives the modified dispersion relation on the lattice
\begin{equation}
	\nabla^2 e^{i\vec k\cdot\vec x} = -k_{\rm eff}^2 e^{i\vec k\cdot\vec x},
\end{equation}
where
\begin{equation}\label{eq:k-eff}
	k_{\rm eff} = \frac{2}{\Delta x}\sqrt{\sum_i\sin^2\frac{k_i\Delta x}{2}} = \frac{2}{\Delta x}\sqrt{\sum_i\sin^2 \frac{\pi n_i}{N}} \ .
\end{equation}
In the continuum limit $k_{\rm eff}$ reduces to $|\vec k|$. This gives the correct expression of momentum magnitude on the lattice and is needed for the correctness of power spectra in the lattice simulation, as pointed out in Ref.~\cite{Caravano:2021pgc}. For example, on the lattice the BD vacuum \eqref{eq:scalar-bd-vacuum} becomes
\begin{equation}\label{eq:scalar-bd-vacuum-lat}
	u_k^{\rm BD} \to u_k^{\rm BD, lat} = \frac{1}{\sqrt{2C_\zeta k_{\rm eff}}} e^{-iC_\zeta k_{\rm eff}\eta}.
\end{equation}

The metric is assumed to be the FLRW metric, \textit{i.e.} we neglect the perturbation of the metric. In the inflation model of Einstein gravity, this approximation is valid since the coupling between the metric and the inflaton is slow-roll suppressed. For GB gravity, this may not hold. To ensure the validity of our lattice computation, we explicitly compute the metric perturbations during our lattice simulation. We find their magnitudes are small ($< 10^{-8}$), thus neglecting metric perturbations is a good approximation. For details, see Appendix \ref{sec:ap:gb-details}.

\subsection{Discrete and continuum power spectra}
Before proceeding let us first derive the relation between continuum and discrete power spectra. For any field $f$, we use the following convention for its Fourier transformation $\tilde f$
\begin{align}
	\tilde f(\vec k) & = \int\frac{\dd^3x}{\sqrt{(2\pi)^3}} e^{-i\vec k\cdot\vec x}f(\vec x) \ ,\\
	f(\vec x) & = \int\frac{\dd^3k}{\sqrt{(2\pi)^3}} e^{i\vec k\cdot\vec x}\tilde f(\vec k) \ ,
\end{align}
and the discretized version on the lattice
\begin{align}
	\hat{\tilde f}(\vec k) & = \sum_{\vec x} e^{-i\vec k\cdot\vec x}f(\vec x),\\
	f(\vec x) & = \frac{1}{N^3}\sum_{\vec k} e^{i\vec k\cdot\vec x}\hat{\tilde f}(\vec k) \ ,
\end{align}
so that we have
\begin{equation}\label{eq:f-hat-f-relation}
    \tilde f = \frac{\Delta x^3}{\sqrt{(2\pi)^3}}\hat{\tilde f} \ .
\end{equation}
The spectrum $\Delta_f$ and its dimensionless version $\mathcal P_f$ \footnote{The definitions of $\Delta_f$ and $\mathcal P_f$ are sometimes interchanged in the literature.} are defined by
\begin{equation}
	\braket{\tilde f(\vec k_1)\tilde f(\vec k_2)} = \Delta_f(k_1) \delta^{(3)}(\vec k_1 + \vec k_2) \ , \ \mathcal P_f = \frac{k^3}{2\pi^2}\Delta_f \ .
\end{equation}
It is easy to verify the convention-independent relation
\begin{equation}
	\braket{f^2(\vec x)} = \int\dd\ln k \mathcal P_f(k).
\end{equation}
On the lattice we are actually working with $\hat{\tilde f}$ and calculating
\begin{equation}
	\braket{\hat{\tilde f}(\vec k_1)\hat{\tilde f}(\vec k_2)} = \tilde\Delta_f(k_1)\delta_{\vec k_1 + \vec k_2, 0} \ .
\end{equation}
Using \eqref{eq:f-hat-f-relation} and the lattice momentum space delta function
\begin{align}
	\delta^{(3)}(\vec k) & = \int\frac{\dd^3x}{(2\pi)^3}e^{-i\vec k\cdot\vec x}\\
	& \to \frac{\Delta x^3}{(2\pi)^3}\sum_{\vec x} e^{-i\vec k\cdot\vec x} = \frac{V}{(2\pi)^3}\delta_{\vec k, 0} \ ,
\end{align}
where $V = L^3 = (N\Delta x)^3$ is the physical volume of the lattice. We then get the relation
\begin{equation}
	\Delta_f(\vec k) = \left(\frac{\Delta x}{N}\right)^3\tilde\Delta_f(\vec k) \ .
\end{equation}
Thus, we finally get
\begin{equation}\label{eq:lat-spectrum-relation}
	\mathcal P_f(k) = \frac{k_{\rm eff}^3}{2\pi^2}\left(\frac{\Delta x}{N}\right)^3\tilde\Delta_f(\vec k) \ .
\end{equation}
Note that we have replaced $k$ with $k_{\rm eff}$.

\subsection{Initial condition}
The initial fluctuation of $\phi$ in our simulation is crucial for calculating the primordial power spectrum. The fluctuation $\delta\phi$ is defined by
\begin{equation}
	\delta\phi = \phi - \braket{\phi}_{\rm lat} \ ,
\end{equation}
where $\braket{.}_{\rm lat}$ denotes the average over lattice points. We need to correctly generate $\phi$ and $\dot\phi$ so that the stochastic correlators $\braket{\delta\phi^2}_{\rm lat}$, $\braket{\delta\phi\delta\dot\phi}_{\rm lat}$ and $\braket{\delta\dot\phi^2}_{\rm lat}$ match the quantum correlators. We then evolve the field using classical equations of motion. At the end of simulation, we calculate the power spectrum using stochastic average
\begin{equation}
	\braket{\delta\phi(\vec x)\delta\phi(\vec y)} = \braket{\delta\phi(\vec x)\delta\phi(\vec y)}_{\rm lat} \ .
\end{equation}
This stochastic approach to simulate quantum systems is known as \ac{twa}, which remains valid when the occupation numbers are large, or quantum fluctuations are small compared to the field magnitude. This condition should hold in the case of inflation.

Perform a mode expansion on $\delta\phi$ and discretize it
\begin{align}
	\delta\phi(\vec x) & = \int\frac{\dd^3k}{\sqrt{(2\pi)^3}}\left(a_{\vec k} \varphi_k + a_{-\vec k}^\dagger\varphi_k^\ast\right) e^{i\vec k\cdot\vec x},\nonumber\\
	& \to \frac{1}{\sqrt V}\sum_{\vec k}\left(\hat a_{\vec k}\varphi_k + \hat a_{-\vec k}^\dagger\varphi_k^\ast\right)e^{i\vec k\cdot\vec x}\label{eq:delta-phi-mode-expand-lat} \ ,
\end{align}
where $\hat a_{\vec k} = \sqrt{(2\pi)^3 / V}a_{\vec k}$ is the discretized ladder operator, satisfying $[\hat a_{\vec k}, \hat a_{\vec k'}^\dagger] = \delta_{\vec k - \vec k'}$. This means we can regard $a_{\vec k}$ as complex Gaussian random numbers with variance $1$. It is easy to verify that this produces correct stochastic correlators. We can then generate $a_{\vec k}$ using the Box–Muller method as described in Ref.~\cite{Caravano:2021pgc}, and compute $\delta\phi$ and $\delta\dot\phi$ using FFT according to \eqref{eq:delta-phi-mode-expand-lat}.

It remains to determine the mode function $\varphi_k$. Lattice simulation is done in spatial flat gauge, where $\delta\phi$ is related to $\zeta$ through the relation $\zeta = -H\delta\phi / \dot\phi$ and
\begin{equation}
	\delta\phi = \phi - \braket{\phi} \ ,
\end{equation}
where $\dot\phi$ should be understood as $\braket{\dot\phi}$. Thus,
\begin{equation}
	\delta\phi = -\frac{\braket{\dot\phi}}{H}\zeta = -\frac{\braket
	{\dot\phi}}{H A_\zeta}\hat u \equiv -\frac{C_\phi}{a}\hat u  \ ,
\end{equation}
so we get the mode function
\begin{equation}
	\varphi_k = -\frac{C_\phi}{a}\hat u_k \ .
\end{equation}
At the start of the simulation we put all modes inside the horizon, so the initial condition \eqref{eq:scalar-bd-vacuum-lat} is used.

\subsection{Extract curvature power spectrum}
To extract curvature power spectrum $\Delta_\zeta$ at the end of lattice simulation, one could use the perturbative relation $\zeta = -H\delta\phi / \braket{\dot\phi}$. However, this does not capture the nonlinear relation between field $\phi$ and $\zeta$. We shall use the method proposed by Ref.~\cite{Caravano:2025diq} to compute $\zeta$. This is a nonperturbative approach based on $\delta N$ formalism. At the end of the lattice simulation, the physical size of each lattice cell has grown larger than the Hubble radius, and the cells effectively become separate universes. The field $\zeta$ then measures the relative count of e-folds required for each lattice site to evolve into the state of constant inflaton field.

Thus, at the end of the lattice simulation, we evolve the fields $a$ and $\phi$ for each lattice site using the background equation of motion until the value of $\phi$ reaches a destination value $\phi_0$. When the field $\phi$ takes the value $\phi_0$, $V_{\rm eff}(\phi, H)$ attains its minimum. We then collect the final value of $a$ of each lattice site into a scalar field $a(\vec x)$ and compute $\zeta$ by
\begin{equation}
	\zeta(\vec x) = \ln a(\vec x) - \braket{\ln a(\vec x)} \ .
\end{equation}
We then proceed to calculate $\Delta_\zeta$ and $\mathcal P_\zeta$ using \eqref{eq:lat-spectrum-relation}.

\section{Numerical results}
\label{sec:results}
We are now ready to carry out the lattice simulation. We first use the numerical method described in Sec.\ref{sec:gb-gravity} to calculate the perturbative primordial power spectrum, then apply the lattice simulation to the frequency range near the peak of the power spectrum. The frequency range of the lattice simulation is controlled by the lattice size $N$ and the physical length $L$. We can see from Eq.\eqref{eq:k-eff} that the minimum and maximum momentum magnitudes are
\begin{equation}
	k_{\rm eff,min} \approx \frac{2\pi}{L} \ , \ k_{\rm eff,max} = \frac{2\sqrt3}{\Delta x} \ ,
\end{equation}
so that $k_{\rm eff,max} / k_{\rm eff,min} = \sqrt3 N / \pi$. The start time of the lattice simulation is chosen so that $k_{\rm eff,min} = 1.3 aH$, where all modes are sub-horizon. We then evolve the fields until $k_{\rm eff,max} = aH / 200$, where all modes are super-horizon. We choose the lattice size to be $N=200$, since the equation of motion for \ac{gb} inflation theory is significantly more complicated than that in the Einstein gravity case, making larger lattice sizes computationally infeasible.

\begin{table}
    \caption{\label{tab:lat-sim-params}Parameter values used.}
	\begin{tabular}{|c|c|c|c|}
	    \hline
		$f$ & $\Lambda$ & $\xi_1$ & $\phi_c$ \\
        \hline
		7.0 & 0.0065 & 9.6 & 30.0\\
		\hline
	\end{tabular}
\end{table}

\begin{figure}
\includegraphics[width=8cm]{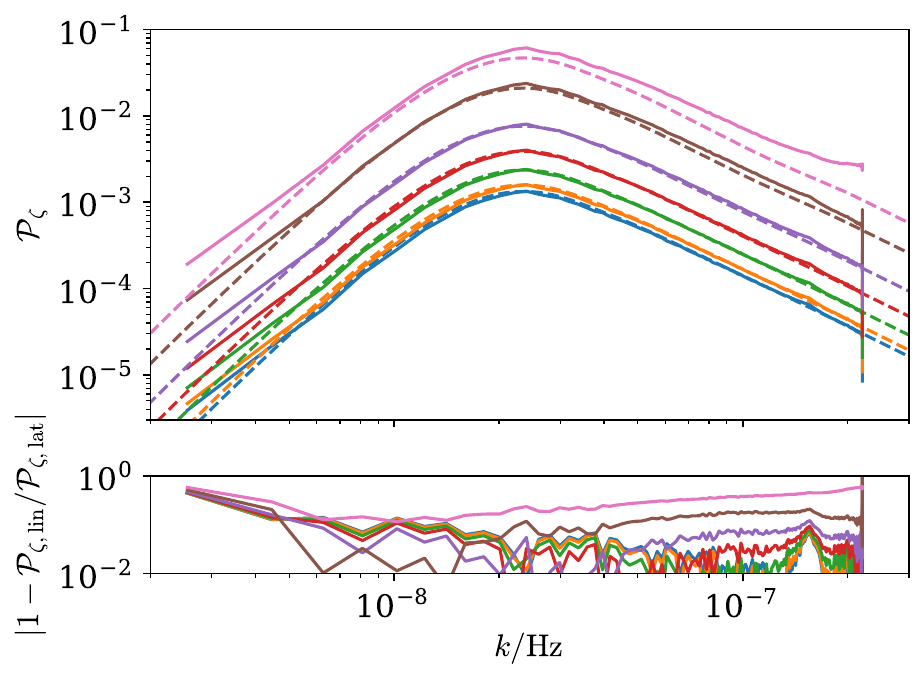}
	\caption{\label{fig:primordial-spectra} Numerical results of primordial curvature spectra. In the upper panel, dashed lines represent the perturbative results $\mathcal P_{\zeta,\rm lin}$, solid lines represent the lattice results $\mathcal P_{\zeta,\rm lat}$. In the lower panel, we plot the relative difference of perturbative and lattice results.}
\end{figure}

We fix the values of parameters $f, \Lambda, \xi_1, \phi_c$ as given in Table.\ref{tab:lat-sim-params} and vary $\alpha$. We select 14 values in $[1.1331\times 10^7, 1.13412\times 10^7]$ with the $i$-th value given by
\begin{equation}\label{eq:alpha}
	\alpha = 1.133 \times 10^7 + a_i \times 10^3 \ , \  (i=0,\cdots, 13) \ ,
\end{equation}
where
\begin{equation}\label{eq:alpha1}
	a_i = \{1, 2, 3, 4, 5, 6, 7, 8, 9, 10, 10.5, 11, 11.1, 11.2\} \ .
\end{equation}
For convenience, we set $\alpha_0=1.133 \times 10^7 + a_0 \times 10^3$. Here, we plot 6 primordial spectra in Fig.\ref{fig:primordial-spectra}. We can see that when the peak of $\mathcal P_\zeta$ reaches $\sim 10^{-2}$, the lattice correction becomes significant.

\section{Scalar induced gravitational waves}\label{sec:SIGW}
\label{sec:observation-constraints}
In the previous sections, we have computed the primordial power spectrum in the \ac{sgb} model using both the conventional perturbative approach and the lattice method. The corresponding numerical results are presented in Fig.~\ref{fig:primordial-spectra}. In this section, we apply these results for the primordial power spectrum to the calculation of second‑order \acp{SIGW} and analyze how the power spectra obtained from the perturbative and lattice approaches affect the resulting energy density spectrum of \acp{SIGW}. Furthermore, by incorporating current \ac{PTA} observational data, we examine the observational constraints on the parameter space of the \ac{sgb} model under both approaches. 

\subsection{Review of \acp{SIGW}}
In the following, we briefly review the main results of \acp{SIGW}. The line element of perturbed \ac{FLRW} spacetime in Newtonian gauge can be expressed as
\cite{Malik:2008im}
\begin{eqnarray}\label{eq:dS}
	\mathrm{d} s^{2}&=&a^{2}\left[-\left(1+2 \phi^{(1)}\right) \mathrm{d} \eta^{2}+\left(\left(1-2 \psi^{(1)}\right) \delta_{i j} \right.\right.\nonumber\\
   &+&\left.\left.\frac{1}{2}h^{(2)}_{ij}\right)\mathrm{d} x^{i} \mathrm{d} x^{j}\right] \ ,
\end{eqnarray}
where $\phi^{(1)}$ and $\psi^{(1)}$ are first-order scalar perturbations, $h^{(2)}_{ij}$ is the second-order tensor perturbation. By substituting Eq.~(\ref{eq:dS}) into the Einstein field equations, retaining only first-order perturbations and simplifying, the equation of motion for the first-order scalar perturbation can be obtained as follows \cite{Ananda:2006af}
\begin{eqnarray}
&&2\mathcal{H} \left(3(c_s^2-w)\mathcal{H}\phi^{(1)}+\phi^{(1)'}+(2+3c_s^2)\psi^{(1)'}   \right)    \label{eq:s1} \nonumber\\
     &&+2\psi^{(1)''}+\Delta \phi^{(1)}-\Delta \psi^{(1)}-2c_s^2\Delta \psi^{(1)}=0 \ ,
\end{eqnarray}
\begin{eqnarray}
  &&\psi^{(1)}-\phi^{(1)}=0 \ . \label{eq:s2}
\end{eqnarray}
 During the \ac{RD} era, $\mathcal{H}=1/\eta$ and $w=c_s^2=1/3$. The analytical solution to Eq.~(\ref{eq:s1}) and Eq.~(\ref{eq:s2}) in momentum space can be written as 
\begin{eqnarray}\label{eq:T1}
    \phi^{(1)}_{\mathbf{k}}(\eta)=\psi^{(1)}_{\mathbf{k}}(\eta)=\frac{2}{3}\zeta_{\mathbf{k}} T_{\phi}(|\mathbf{k}| \eta) \ ,
\end{eqnarray}
where $\zeta_{\mathbf{k}}$ is the primordial curvature perturbation. The transfer function $T_\phi(x)$ can be expressed as \cite{Baumann:2007zm}
\begin{eqnarray}\label{eq:TRD}
    T_{\phi}(x)=\frac{9}{x^{2}}\left(\frac{\sqrt{3}}{x} \sin \left(\frac{x}{\sqrt{3}}\right)-\cos \left(\frac{x}{\sqrt{3}}\right)\right) \ ,
\end{eqnarray}
where $x=|\mathbf{k}|\eta$. After solving the first-order scalar perturbation, we substitute Eq.~(\ref{eq:dS}) into the Einstein field equations and retain only the transverse–traceless part of the second-order perturbation. In this way, the equation of motion for the second-order tensor perturbation is obtained as
\begin{equation}\label{eq:h}
	\begin{aligned}
		h_{ij}^{(2)''}(\eta,\mathbf{x})&+2 \mathcal{H} h_{ij}^{(2)'}(\eta,\mathbf{x})-\Delta h_{ij}^{(2)}(\eta,\mathbf{x}) \\
        &=-4 \Lambda_{ij}^{lm} \mathcal{S}^{(2)}_{lm}(\eta,\mathbf{x})  \ ,
	\end{aligned}
\end{equation}
where $\Lambda_{ij}^{lm}$ is the decomposed operator to extract the transverse and traceless terms \cite{Chang:2020tji}. The source term in Eq.~(\ref{eq:h}) is given by
\begin{eqnarray}
	\mathcal{S}^{(2)}_{lm}(\eta,\mathbf{x})&=&\partial_{l} \phi^{(1)} \partial_{m} \phi^{(1)} +4 \phi^{(1)} \partial_{l} \partial_{m} \phi^{(1)}\nonumber\\
    &-&\frac{1}{ \mathcal{H}}\left(\partial_{l} \phi^{(1)'} \partial_{m} \phi^{(1)}+\partial_{l} \phi^{(1)}  \partial_{m} \phi^{(1)'}\right) \nonumber\\
	&-&\frac{1}{ \mathcal{H}^{2}} \partial_{l} \phi^{(1)'} \partial_{m}  \phi^{(1)'} \ .
	\label{eq:1} 
\end{eqnarray}
Eq.~(\ref{eq:h}) and Eq.~(\ref{eq:1}) arise directly from the second-order perturbation of the Einstein field equations in \ac{FLRW} background. These relations indicate that the first-order scalar perturbation couples to the second-order tensor perturbation via the second-order cosmological perturbation equations. If the primordial curvature perturbation on small scales is sufficiently enhanced, it affects the corresponding first-order scalar perturbation and, via Eq.~(\ref{eq:h}) and Eq.~(\ref{eq:1}), leads to the generation of second-order tensor modes with observable impact. Gravitational waves produced through this mechanism are referred to as second-order \acp{SIGW}.

The solution of the second-order \acp{SIGW} can be expressed in the following form
\begin{equation}\label{eq:hms}
\begin{aligned}
h_{i j}^{(2)}(\mathbf{x}, \eta)=\int \frac{\mathrm{d}^3 k}{(2 \pi)^{3 / 2}} e^{i \mathbf{k} \cdot \mathbf{x}}\sum_{\lambda}h_{\mathbf{k}}^{\lambda,(2)}(\eta) \varepsilon_{i j}^{\lambda}(\mathbf{k}) \  ,
\end{aligned}
\end{equation}
where $\varepsilon_{i j}^{\lambda}(\mathbf{k})$$(\lambda=+,\times)$ are polarization tensors. The Fourier components $ h^{\lambda,(2)}_{\mathbf{k}}(\eta)$$(\lambda=+,\times)$ of \acp{SIGW} can be expressed as
\begin{eqnarray}
	h^{\lambda,(2)}(\eta,\mathbf{k})&=&\frac{4}{9}\int\frac{\mathrm{d}^3p}{(2\pi)^{3/2}}\varepsilon^{\lambda,lm}\left(\mathbf{k}\right)p_lp_m \nonumber\\
    &&I^{(2)}\left( u,v,x \right)\zeta_{\mathbf{k}-\mathbf{p}}\zeta_{\mathbf{p}} \ , \label{eq:h11}
\end{eqnarray}
where
\begin{eqnarray}
        I^{(2)}&=&
        \frac{27 (u^2 + v^2 - 3)}{k^2u^3 v^3 x}
        \bigg(
            \sin x 
            \left(
                -4 u v + (u^2  \right.\nonumber\\
            &+&\left.v^2- 3) 
                \ln \left| \frac{3 - (u + v)^2}{3 - (u - v)^2} \right|
            \right)-\pi(u^2 \nonumber\\
        &+& v^2 - 3) \Theta(v + u - \sqrt{3}) \cos x
        \bigg)  \label{eq:11I}
\end{eqnarray}
is the kernel function \cite{Kohri:2018awv}. With the analytical solution of the second-order \acp{SIGW} obtained, we can compute the corresponding two-point correlation function and derive the energy density spectrum of the second-order \acp{SIGW} as follows \cite{Espinosa:2018eve,Kohri:2018awv}
\begin{align}
    \Omega_{\mathrm{GW}}^{(2)}(k) &= \int_{0}^{\infty} \mathrm{d}v \int_{|1-v|}^{1+v} \mathrm{d}u\,
    \mathcal{P}_{\zeta}(uk) \mathcal{P}_{\zeta}(vk) \notag \\
    &\times  \frac{3}{1024 u^8 v^8} (u^2 + v^2 - 3)^2 \notag \\
    &\times \left[4v^2 - (1 + v^2 - u^2)^2\right]^2 \notag \\
    &\times \left\{ \left[(u^2 + v^2 - 3) \ln \left| \frac{3 - (u + v)^2}{3 - (u - v)^2} \right| - 4uv \right]^2 \right. \notag \\
    & + \left.  \pi^2 (u^2 + v^2 - 3)^2 \Theta \Big(u + v - \sqrt{3} \Big) \right\} \ .
    \label{eq:1Oss}
\end{align}
At this stage, all computational details of the second-order \acp{SIGW} are encapsulated in Eq.~(\ref{eq:1Oss}). Once a specific primordial power spectrum $\mathcal{P}_{\zeta}(k)$ is provided and substituted into Eq.~(\ref{eq:1Oss}), the energy density spectrum of the second-order \acp{SIGW} can be calculated. Furthermore, Eq.~(\ref{eq:1Oss}) provides the energy density spectrum of \acp{SIGW} during the \ac{RD} era. Taking into account the thermal history of the universe, the current total energy density spectrum $\bar{\Omega}^{(2)}_{\mathrm{GW,0}}(k)$ is given by
\begin{equation}
\bar{\Omega}^{(2)}_{\mathrm{GW,0}}(k) = \Omega_{\mathrm{rad},0}\left(\frac{g_{*,\rho,e}}{g_{*,\rho,0}}\right)\left(\frac{g_{*,s,0}}{g_{*,s,e}}\right)^{4/3}\bar{\Omega}^{(2)}_{\mathrm{GW}}(k) \ ,
\end{equation}
where $\Omega_{\mathrm{rad},0}$ ($ =4.2\times 10^{-5}h^{-2}$) is the energy density fraction of radiation today.

\begin{figure}
\includegraphics[width=8cm]{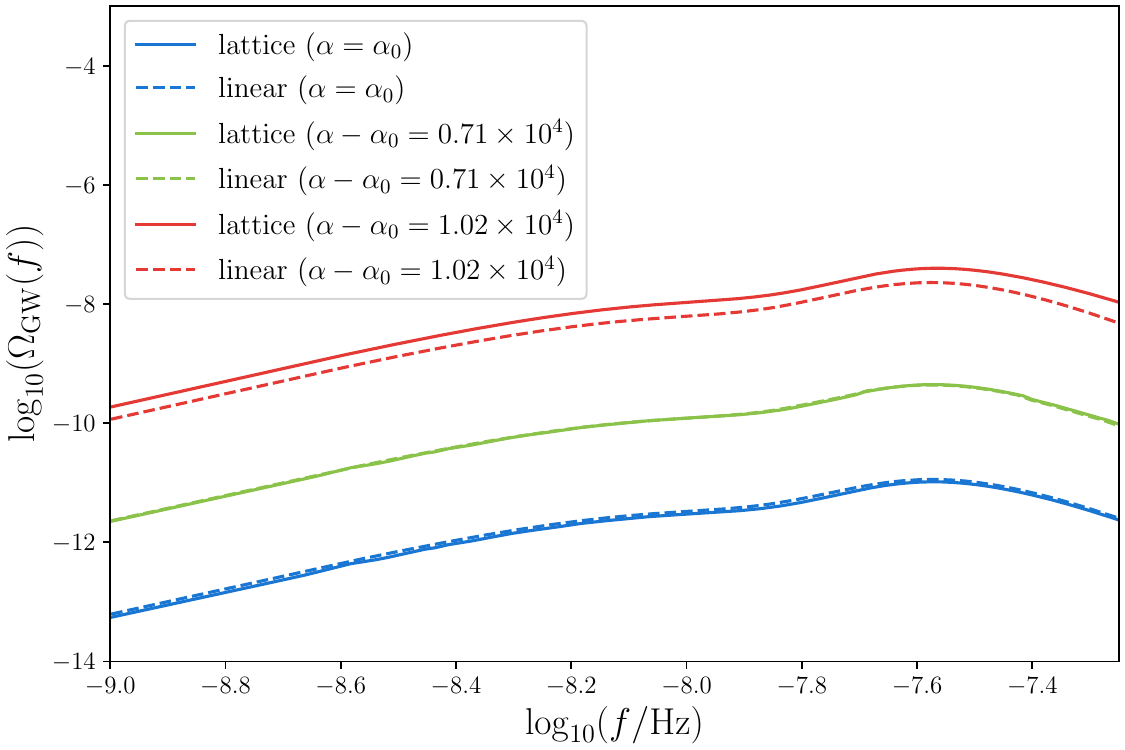}
	\caption{\label{fig:spectrum} Energy density spectra of second-order \acp{SIGW} computed using the primordial power spectra obtained from the lattice method and the perturbative method. The solid and dashed curves correspond to the lattice and perturbative results, respectively. The physical meaning of the parameter $\alpha$ is provided by Eq.~(\ref{eq:alpha}). }
\end{figure}

In Fig.~\ref{fig:spectrum}, we present the energy density spectra of \acp{SIGW} corresponding to the primordial power spectra obtained from both the perturbative method and the lattice method. When the primordial spectrum exhibits a large amplitude, the impact of lattice inflation on the energy density spectra becomes clearly significant.

\subsection{\ac{PTA} observations}
In the previous subsection, we reviewed the main results of \acp{SIGW} and computed the corresponding energy density spectrum. We now discuss the impact of the \ac{sgb} model, as calculated using the lattice method, on \ac{PTA} observations. More precisely, we construct the likelihood using the \ac{KDE} representations of the free spectra~\cite{Mitridate:2023oar,Lamb:2023jls}
\begin{equation}
    \ln \mathcal{L}(d|\theta)=\sum_{i=1}^{N_f} p(\Phi_i,\theta)\, .
\end{equation}
Here, $p(\Phi_i,\theta)$ denotes the probability of observing $\Phi_i$ given the parameter set $\theta$, and $\Phi_i=\Phi(f_i)$ is the time delay,
\begin{equation}
\Phi(f)=\sqrt{\frac{H_0^2\,\Omega_{\mathrm{GW}}(f)}{8\pi^2 f^5 T_{\mathrm{obs}}}}\, ,
\end{equation}
with $H_0=h\times100\,\mathrm{km\,s^{-1}\,Mpc^{-1}}$ the present Hubble constant. In our analysis, we adopt the \ac{KDE} representation of the first 14 frequency bins of the HD-correlated free spectrum from the NANOGrav 15-year dataset~\cite{Nanograv:KDE}. Bayesian inference is carried out using \textsc{bilby}~\cite{bilby_paper} with the \textsc{dynesty} nested sampler~\cite{Speagle:2019ivv,dynesty_software}.

Furthermore, to rigorously evaluate the viability of different scenarios in explaining the current \ac{PTA} observations, we also consider a mixed case in which both \acp{SMBHB} and \acp{SIGW} contribute to the observed signal. The energy density spectrum of \acp{SMBHB} is given by~\cite{Mitridate:2023oar,NANOGrav:2023hvm}
\begin{equation}
    \Omega_{\mathrm{GW}}^{\mathrm{BH}}(f)
    = \frac{2\pi^2 A_{\mathrm{BHB}}^2}{3H_0^2 h^2}
      \left(\frac{f}{\mathrm{year}^{-1}}\right)^{5-\gamma_{\mathrm{BHB}}}
      \mathrm{year}^{-2}\, .
\end{equation}
The prior distribution for $(\log_{10}A_{\mathrm{BHB}},\gamma_{\mathrm{BHB}})$ is modeled as a multivariate normal distribution~\cite{NANOGrav:2023hvm} with mean and covariance
\begin{equation}
\begin{aligned}
    \boldsymbol{\mu}_{\mathrm{BHB}} &= 
    \begin{pmatrix} -15.6 \\ 4.7 \end{pmatrix}, \\
    \boldsymbol{\sigma}_{\mathrm{BHB}} &= 
    0.1 \times
    \begin{pmatrix}
        2.8 & -0.026 \\
        -0.026 & 2.8
    \end{pmatrix}.
\end{aligned}
\end{equation}

\begin{figure}
\includegraphics[width=8cm]{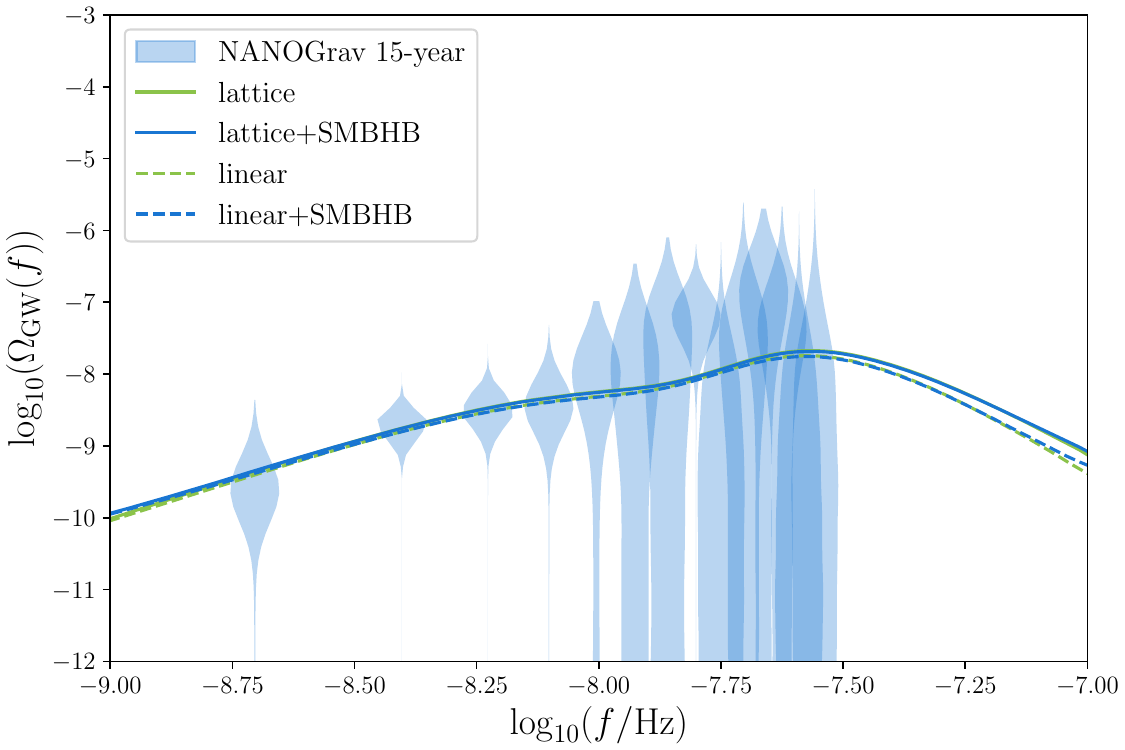}
	\caption{\label{fig:violinplot} Energy density spectra of second-order \acp{SIGW}. The solid and dashed curves correspond to the lattice and perturbative results, respectively. The energy density spectra derived from the free spectrum of the NANOGrav 15-year dataset is shown with blue shading. }
\end{figure}

\begin{figure}
\includegraphics[width=8cm]{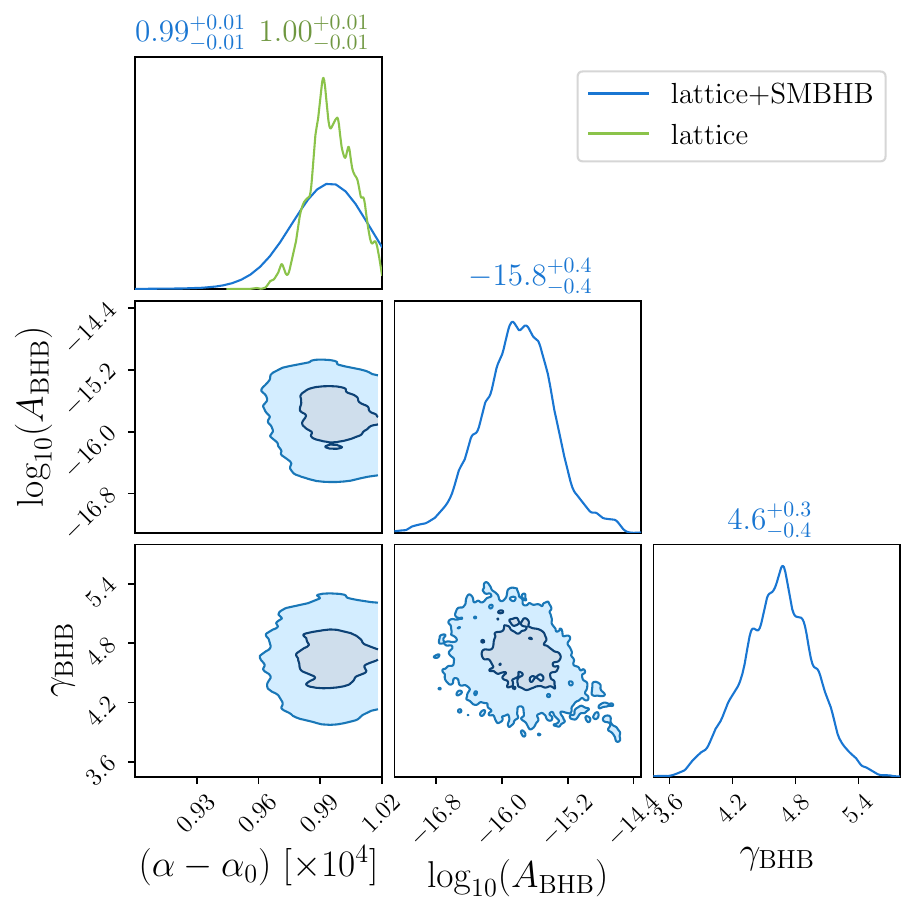}
	\caption{\label{fig:CLa} The corner plot of the posterior distributions. The contours in the off-diagonal panels denote the $68\% $ and $95 \%$ credible intervals of the 2D posteriors. The numbers above the figures represent the median values and $1$-$\sigma$ ranges of the parameters. The green and blue curves in the figure correspond to the results obtained from the lattice method and from the lattice method combined with the \ac{SMBHB} contribution, respectively.}
\end{figure}

\begin{figure}
\includegraphics[width=8cm]{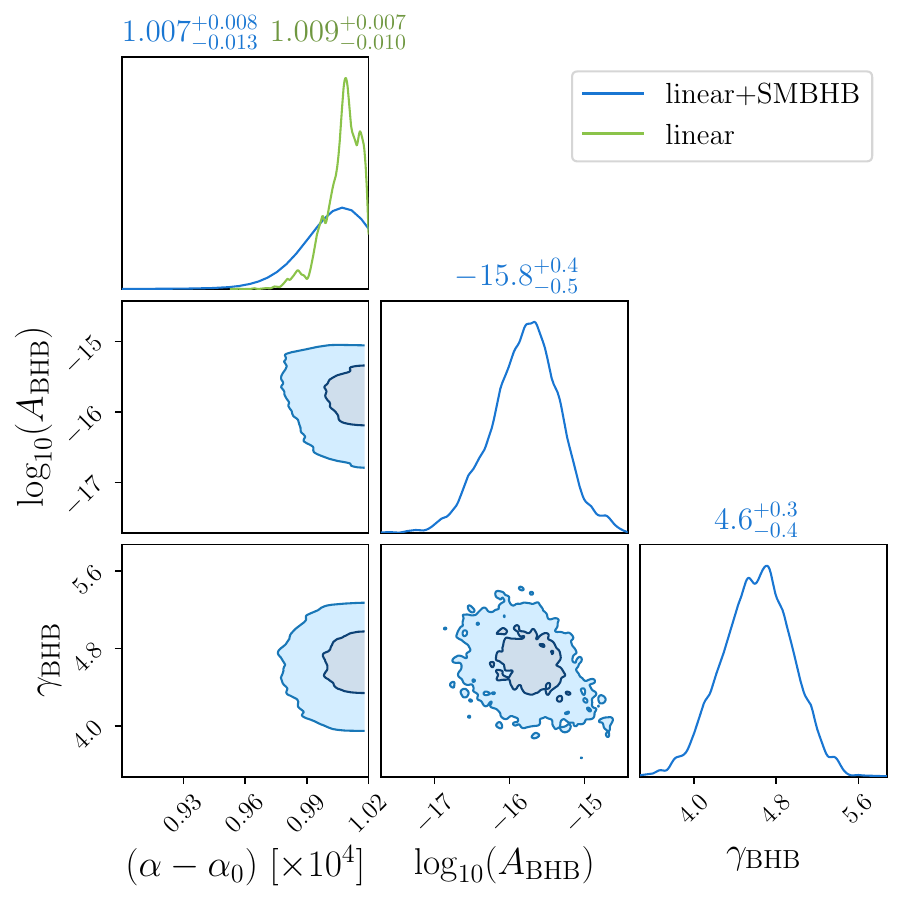}
	\caption{\label{fig:CLn} The corner plot of the posterior distributions. The contours in the off-diagonal panels denote the $68\% $ and $95 \%$ credible intervals of the 2D posteriors. The numbers above the figures represent the median values and $1$-$\sigma$ ranges of the parameters. The green and blue curves in the figure correspond to the results obtained from the linear‑perturbation method and from the linear‑perturbation method supplemented by the \ac{SMBHB} contribution, respectively.}
\end{figure}

Figure~\ref{fig:violinplot} illustrates the energy density spectra associated with second-order \acp{SIGW}. The posterior distributions constrained by the \ac{PTA} data are presented in Fig.~\ref{fig:CLa} and Fig.~\ref{fig:CLn}. Since the lattice corrections coincide with the perturbative results when the parameter $\alpha-\alpha_0$ in Eq.~(\ref{eq:alpha}) is small, we adopt uniform priors spanning the range $[9000,10200]$. By comparing Fig.~\ref{fig:CLa} and Fig.~\ref{fig:CLn}, we find that lattice corrections affect only the posterior distribution in the region with large values of the parameter $\alpha-\alpha_0$, consistent with the results shown in Fig.~\ref{fig:spectrum}. Specifically, the amplitude of the primordial power spectrum increases as $\alpha-\alpha_0$ becomes larger. When $\alpha-\alpha_0$ is small, the amplitude of the primordial spectrum remains low, and the lattice and perturbative results are essentially identical; hence, lattice corrections do not influence either the energy density spectrum or the posterior distribution of $\alpha-\alpha_0$. As $\alpha-\alpha_0$ increases, the amplitude of the primordial spectrum grows accordingly, and the impact of lattice corrections becomes progressively more significant, thereby affecting the posterior distribution of $\alpha-\alpha_0$.


\begin{figure}
\includegraphics[width=8cm]{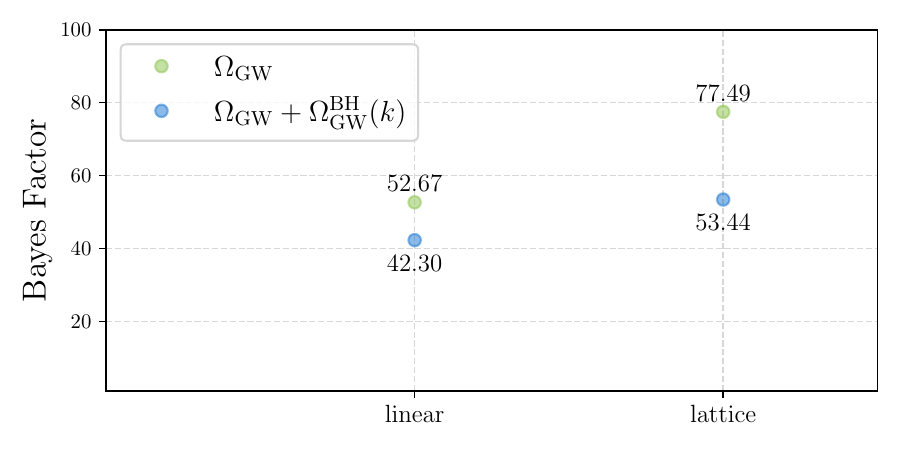}
	\caption{\label{fig:bayes} The Bayes factors between different models. The vertical axis represents the Bayes factor of different models relative to \ac{SMBHB}, and the horizontal axis represents the different models. The results indicate that lattice corrections enhance the Bayes factor of the \ac{sgb} model.}
\end{figure}

Furthermore, to compare competing models, we compute the Bayes factor, defined as $B_{i,j}=Z_i/Z_j$, where $Z_i$ denotes the Bayesian evidence of model $H_i$. As shown in Fig.~\ref{fig:bayes}, we compute the Bayes factors for different models. After incorporating the lattice corrections, the Bayes factor of the \ac{sgb} model becomes larger. Specifically, as illustrated in Fig.~\ref{fig:CLa} and Fig.~\ref{fig:CLn}, the peak of the posterior distribution obtained from the lattice method (green curve in Fig.~\ref{fig:CLa}) is shifted leftward compared with that in  Fig.~\ref{fig:CLn}. Consequently, within the parameter range we consider, the lattice method allows a larger portion of the parameter values of $\alpha-\alpha_0$ to yield energy density spectra that fit the current \ac{PTA} observations. This result indicates that lattice corrections enhance the ability of the \ac{sgb} model to dominate the \ac{PTA} observations.

\section{Conclusion}
\label{sec:conclusion}
We investigated the single-field inflation in \ac{sgb} gravity using lattice simulations. We focus on the \ac{usr} scenario in which the primordial curvature perturbation is significantly enhanced on small scales, with the corresponding peak frequency lying in the \ac{PTA} band. Since sufficiently large primordial perturbations can lead to significant higher-order corrections and potentially invalidate the perturbative treatment, we have compared the primordial power spectra obtained from the conventional perturbative approach and the lattice method, and further examined the implications for second-order \acp{SIGW} and current \ac{PTA} observations. More precisely, we find that the lattice results are \textit{larger} than perturbative results, and the difference between lattice and perturbative predictions becomes increasingly pronounced as the amplitude of the primordial power spectrum grows. Since the energy density spectrum of \acp{SIGW} is proportional to the square of the amplitude of the primordial curvature power spectrum, a primordial power spectrum with an amplitude of order $10^{-2}$ can significantly modify the resulting \ac{SIGW} signal. This demonstrates that non‑perturbative corrections during inflation can have observable consequences at later cosmological epochs through the \acp{SIGW} channel.

In addition, by incorporating current \ac{PTA} observational data, we analyze the parameter space of different \ac{SGWB} models. Specifically, we considered both the \acp{SIGW}-dominated scenario and the mixed scenario in which \acp{SIGW} and \ac{SMBHB} contribute to the \ac{SGWB}. The Bayesian analysis shows that lattice corrections mainly affect the posterior distribution in the parameter region where the primordial power spectrum has a large amplitude, while the lattice and perturbative results remain nearly identical for smaller amplitudes. Moreover, within the parameter range considered in this work, the inclusion of lattice corrections increases the Bayes factors for the \ac{sgb} model relative to the \ac{SMBHB} scenario. In particular, the lattice calculation allows a larger fraction of the \ac{sgb} parameter space to produce energy density spectra of \acp{SIGW} compatible with the current \ac{PTA} observations. Our results highlight the importance of non‑perturbative effects in inflation models with strongly enhanced small-scale primordial perturbations. A fully non‑perturbative treatment is consequently important when the primordial power spectrum approaches the regime where perturbation theory becomes unreliable. The lattice approach provides a useful framework for studying this regime and offers a more complete connection between inflationary dynamics and \ac{SGWB}.

\appendix

\section{Details of scalar Gauss-Bonnet theory}
\label{sec:ap:gb-details}
The full equation of motion of \ac{sgb} theory is given by \eqref{eq:gb-phi-eom}, \eqref{eq:gb-a-eom}, where $\braket{\cdots}$ denotes spatial average.
\begin{widetext}
    \begin{align}
        \ddot{\phi} \times M &= -64a^6V'(\phi)+24H^2a^6\left(5\dot{\phi}^{\,2}-2V(\phi)\right)\xi'(\phi)+16a^2(\Box\phi)\left(4a^2-2Ha^2\dot{\phi}\,\xi'(\phi)\right)
        \nonumber\\
        &\quad+32Ha^6\dot{\phi}\left(-6+ V'(\phi)\xi'(\phi)\right)-24H^4a^6\xi'(\phi)\left(-2+\dot{\phi}^{\,2}\xi''(\phi)\right)\label{eq:gb-phi-eom}\\
        \ddot a \times M & = \Big\langle 16a^6\left(\dot{\phi}^{\,2}-2V(\phi)\right)+48H^3a^6\dot{\phi}\,\xi'(\phi)
        \nonumber\\
        &\quad+16H^2a^4\left[-(\Box\phi)\xi'(\phi)+a^2\left(2+ V'(\phi)\xi'(\phi)-\dot{\phi}^{\,2}\xi''(\phi)\right)\right]\Big\rangle\label{eq:gb-a-eom}\\
        M & = 4a^2\left(16a^4+H a\left(-8a^3\dot{\phi}\,\xi'(\phi)+6H^3a^3\left[\xi'(\phi)\right]^2\right)\right)
    \end{align}
\end{widetext}
Now we examine the metric perturbation. Consider the spatially flat gauge with
\begin{equation}
    \delta\phi \ne 0, \qquad \delta g_{\mu\nu} = \begin{pmatrix}
        -2A & \partial_i B\\
        \partial_i B & 0_{3 \times 3}
    \end{pmatrix}.
\end{equation}
After a straightforward but lengthy calculation, the equation of motion for $A$ and $\Box B$ are found to be given by \eqref{eq:gb-pert-a-eom} and \eqref{eq:gb-pert-b-eom} respectively. We can then check the magnitudes of $A$ and $\Box B$ by evaluating \eqref{eq:gb-pert-a-eom} and \eqref{eq:gb-pert-b-eom} during the lattice simulation. We find that, even for the case with largest amplitude, \textit{i.e.} $\alpha = 1.13412 \times 10^7$, we have $|A| \leqslant 10^{-6}$ and $|\Box B / a^2| \leqslant 5 \times 10^{-9}$, which confirms that neglecting metric perturbations is a good approximation.
\begin{widetext}
    \begin{align}
        A \times M_2 & = H^2\left(-H\delta\phi+\dot{\delta\phi}\right)\xi'(\phi)+\delta\phi\,\dot{\phi}\left(-2+H^2\xi''(\phi)\right),\label{eq:gb-pert-a-eom}\\
        \Box B \times M_2^2 & = -8H\,\delta\phi\,a^2V'(\phi)+4H^3\left(-3\delta\phi\,H^2a^2+\Box\delta\phi\right)\xi'(\phi)+H\dot{\phi}\Bigg[-a^2\dot{\delta\phi}\left(8+3H^4\left[\xi'(\phi)\right]^2\right)
        \nonumber\\
        &\qquad\qquad+3H\Bigg(-8\delta\phi\,a^2+2\delta\phi\,a^2 V'(\phi)\xi'(\phi)+H^2
        \left(4\delta\phi\,H^2a^2-\Box\delta\phi\right)\left[\xi'(\phi)\right]^2\Bigg)\Bigg]
        \nonumber\\
        &\quad-2\delta\phi\,a^2\dot{\phi}^{\,3}\left(-2+H^2\xi''(\phi)\right)+H^2a^2\dot{\phi}^{\,2}\xi'(\phi)\left(4\dot{\delta\phi}+\delta\phi\,H\left(26-3H^2\xi''(\phi)\right)\right).\label{eq:gb-pert-b-eom}\\
        M_2 & = H (3H\dot\phi\xi'(\phi) - 4).
    \end{align}
\end{widetext}

\bibliography{refs}

\end{document}